\documentclass[12pt]{article}
\usepackage{epsfig}
\usepackage{pstricks,rotating, graphicx}
\usepackage{a4}
\usepackage{latexsym}
\usepackage{cite}

\usepackage{color}
\usepackage{colordvi}
\usepackage[hang]{subfigure}

\graphicspath{{Pics/}}
\DeclareGraphicsExtensions{.eps,.ps}

\usepackage[width=16.5cm, left=2.2cm,top=2.5cm,height=24.cm]{geometry}

\usepackage{pslatex}
\usepackage{txfonts}
\usepackage[latin1]{inputenc}
\usepackage[T1]{fontenc}

\newcommand{\GeV}{\ensuremath{\,\mathrm{GeV}}}
\newcommand{\TeV}{\ensuremath{\,\mathrm{TeV}}}

\def\alphas{{\alpha_s}}

\def\MSbar{{$\overline{\mbox{MS}}\,$}}

\begin{document}

\begin{titlepage}
\noindent
DESY 11-001 \\
SFB/CPP-11-01 \\
LPN 11-01 
\vspace{1.3cm}

\begin{center}
  {\bf
    \Large
    Higher order constraints on the Higgs production rate \\ from fixed-target DIS data \\
  }
  \vspace{1.5cm}
  {\large
    S.~Alekhin$^{\, a,b}$\footnote{{\bf e-mail}: sergey.alekhin@ihep.ru},
    J.~Bl\"umlein$^{\, a}$\footnote{{\bf e-mail}: johannes.bluemlein@desy.de}
    and S.~Moch$^{\, a,}$\footnote{{\bf e-mail}: sven-olaf.moch@desy.de} \\
  }
  \vspace{1.2cm}
  {\it
    $^a$Deutsches Elektronensynchrotron DESY \\
    Platanenallee 6, D--15738 Zeuthen, Germany \\
    \vspace{0.2cm}
    $^b$Institute for High Energy Physics \\
    142281 Protvino, Moscow region, Russia\\
  }
  \vspace{1.4cm}
  \large {\bf Abstract}
  \vspace{-0.2cm}
\end{center}
The constraints of fixed-target DIS data in fits of parton distributions including QCD
corrections to next-to-next-to leading order are studied. 
We point out a potential problem in the analysis of the NMC data 
which can lead to inconsistencies in the extracted value for $\alphas(M_Z)$ 
and the gluon distribution at higher orders in QCD. 
The implications for predictions of rates for Standard Model Higgs boson production at hadron colliders are investigated.
We conclude that the current range of excluded Higgs boson masses at the Tevatron appears to be much too large.
\end{titlepage}
\setcounter{footnote}{0}

The Higgs boson is the last missing cornerstone of the Standard Model (SM).
Searches for the Higgs boson are in the very center of the experimental
activity at the current hadron colliders.
At the moment, from the combined data of the Tevatron experiments exclusion
limits for the SM Higgs boson are derived in a certain mass range~\cite{higgs@tev:2010ar}, 
while the LHC experiments are in the process of improving their discovery (or exclusion) potential 
with increasing integrated luminosity.
At the Tevatron and the LHC the Higgs boson can be produced in a large variety of channels, 
with the gluon-gluon fusion process dominating by roughly one order of magnitude over 
vector-boson fusion or Higgs-Strahlung. 
Precision predictions for the respective production rates are a key ingredient in the experimental searches, 
the higher-order radiative corrections usually being known 
to next-to-next-to-leading order (NNLO) in QCD and to next-to-leading order (NLO) 
as far as electro-weak corrections are concerned~(see e.g.~\cite{Djouadi:2005gi,Harlander:2007zz}).
As favorable features, predictions based on higher-order quantum corrections display 
an apparent convergence of the perturbative expansion and a substantially 
reduced dependence on the choice of the factorization and renormalization scales.
For the particular case of Higgs boson production in gluon-gluon fusion even 
the NNLO corrections in QCD are still sizable, e.g. roughly 30\% for the total cross section, 
so that NNLO accuracy~\cite{Harlander:2002wh,Anastasiou:2002yz,Ravindran:2003um} is mandatory.

Phenomenology at hadron colliders, however, also has to address 
the uncertainty due to the non-perturbative parameters, such as the parton distributions (PDFs), 
the value of the strong coupling constant $\alphas(M_Z)$ and the mass $m$ of the heavy quarks. 
It has become obvious, that currently the largest differences between the various predictions of the Higgs boson cross sections 
at Tevatron and the LHC are of precisely this origin~\cite{Baglio:2010um,Alekhin:2010dd}.
In this Letter we investigate this point in detail.
We are concerned here with the value of the strong coupling constant
$\alphas(M_Z)$ and the PDFs as determined in global fits of PDFs 
and we would like to pin down the source of the resulting differences 
between the PDF sets of ABKM~\cite{Alekhin:2009ni,Alekhin:2010iu} and others.
The issue of precision input for the value for heavy-quark masses $m$ 
has recently been solved by using the running mass in the \MSbar scheme~\cite{Alekhin:2010sv}. 

PDFs as determined in global fits rely on a variety of data predominantly from deep-inelastic scattering (DIS) experiments 
in order to cover the entire kinematic range in the parton momentum fractions $x_p$.
Global PDF fits also combine scattering data with different beams and different targets 
to allow for the separation of the individual quark flavors.
Current Higgs boson searches probe the PDFs at scales $\mu$ 
of the order of the typical values of the Higgs boson mass $M_H$, e.g. say $\mu = 165~\GeV$, 
and in an effective $x_p$-range determined by $\langle x_p \rangle = M_H/\sqrt{s}$, 
where $\sqrt{s}$ is the center-of-mass energy of the collider.
The production region at  Tevatron is governed  by average values of $\langle x_p \rangle \sim 0.1$, 
while those at e.g. $\langle x_p \rangle \sim 0.03$ are characteristic for the LHC at $\sqrt{s} = 7~\TeV$. 
In this $x_p$-range the relevant experimental constraints on the PDFs are to a
great extent due to DIS fixed-target experiments 
(BCDMS~\cite{Benvenuti:1989fm,Benvenuti:1989rh}, SLAC~\cite{Whitlow:1991uw}, 
NMC~\cite{Arneodo:1996qe}, etc).
Thus, the processing of these data in global fits as well as any assumptions being made 
must come under scrutiny.
As a matter of fact, as will be shown below, 
differences in the treatment of higher-order radiative corrections to fixed-target DIS data 
can be made responsible for the bulk of the deviations in cross section predictions 
based either on the PDF set ABKM~\cite{Alekhin:2009ni,Alekhin:2010iu} or 
on MSTW~\cite{Martin:2009iq}, the latter being the basis of the current Tevatron Higgs searches~\cite{higgs@tev:2010ar}. 

\bigskip

The fixed-target data are typically provided as differential cross sections of charged-lepton DIS off nucleons.
In the neutral current case the latter can be written in the one-photon exchange 
approximation as,
\begin{eqnarray}
\label{eq:sigma}
\frac{d^2\, \sigma(x,Q^2)}{dx dQ^2} &=& 
\frac{4 \pi \alpha^2}{x Q^4} \, 
\Biggl\{ 1 - y - x y \frac{M^2}{s} + \left(1-\frac{2 m_l^2}{Q^2}\right) 
\left(1+4 x^2 \frac{M^2}{Q^2}\right) \frac{y^2}{2(1+R(x,Q^2))} 
\Biggr\} \, F_2(x,Q^2) 
\, ,
\end{eqnarray}
where $\alpha$ is the fine structure constant, $Q^2$ the (space-like) four-momentum transfer squared, 
$M$ the proton mass and $m_l$ the mass of the incident charged lepton. 
The Bjorken scaling variable is denoted by $x$ and the inelasticity as $y$ (see e.g.~\cite{Arbuzov:1995id}).
The differential cross section in Eq.~(\ref{eq:sigma}) depends on the DIS structure functions $F_2$ and $F_L$. 
The dependence on $F_L$ can also be parametrized by the 
ratio of the longitudinally to transversely polarized virtual
photon absorption cross sections, $R=\sigma_L/\sigma_T$.
The perturbative expansion for the DIS structure functions $F_2$ and $F_L$ in QCD reads 
\begin{eqnarray}
  \label{eq:F2FL}
  F_2 =
  \sum_{l=0}^\infty \alphas^l F_2^{(l)}
\, ,
\qquad\qquad
  F_L =
  \alphas \, \sum_{l=0}^\infty \alphas^l F_L^{(l)}
\, ,
\end{eqnarray}
with the higher-order corrections being known to NNLO for the PDF evolution~\cite{Moch:2004pa,Vogt:2004mw}, 
as well as for the Wilson coefficients of $F_2$~\cite{vanNeerven:1991nn,Zijlstra:1991qc,Zijlstra:1992qd} 
and $F_L$~\cite{SanchezGuillen:1990iq,Moch:2004xu} (see also \cite{Vermaseren:2005qc} and Refs. therein).
Since we investigate Higgs boson production to NNLO in QCD, a consistent
treatment of the PDFs and of the fixed-target DIS data therefore 
also requires the NNLO corrections~\footnote{%
Note that the correct ${\cal O}(\alphas^2)$ result for $F_L$ was only available 
after the final publication of the BCDMS data~\cite{Benvenuti:1989fm,Benvenuti:1989rh}, 
but before the NMC analysis~\cite{Arneodo:1996qe}.} 
for both $F_2$ and $F_L$.

\bigskip

\begin{figure}[t!]
\centering
\vspace*{10mm}
    {
    \includegraphics[angle=0,width=8.0cm]{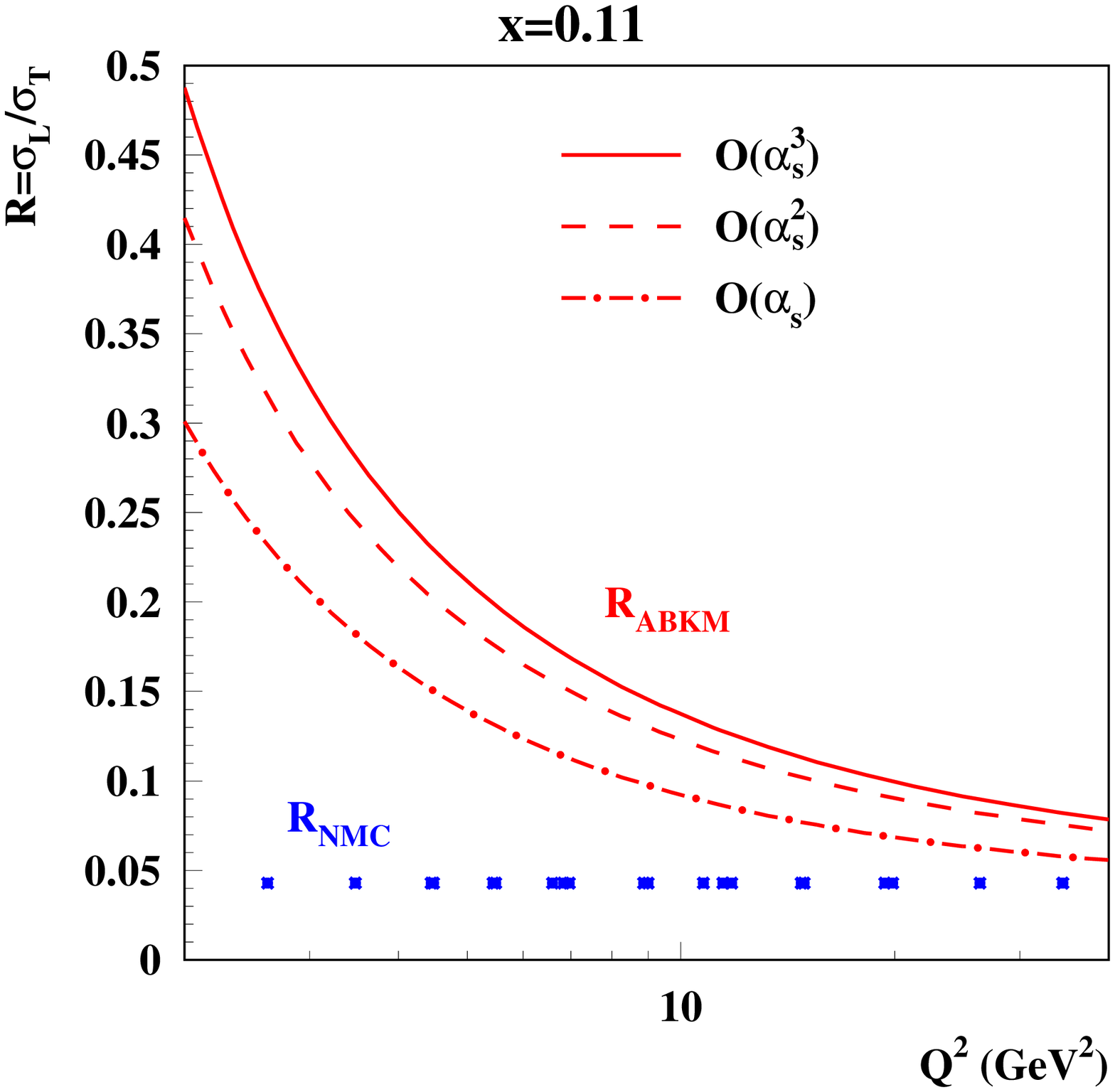}
    \includegraphics[angle=0,width=8.0cm]{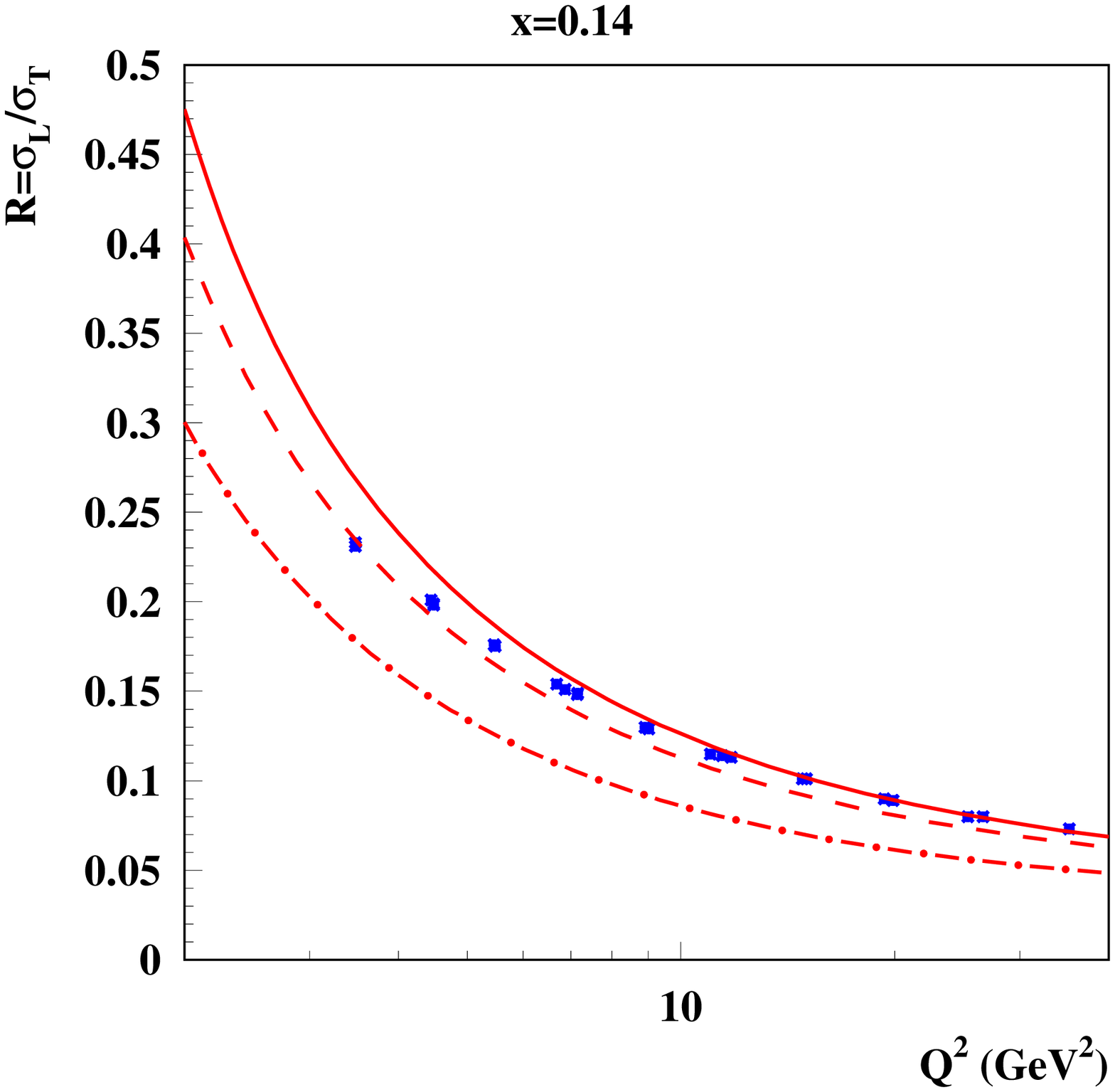}
    }
    \caption{ \small
      \label{fig:R-nmc}
      The ratio of cross sections $R=\sigma_L/\sigma_T$ 
      for longitudinally to transversely polarized virtual photon DIS 
      as a function of $Q^2$ for different values of $x$.
      $R_{\rm NMC}$ denotes the results of~\cite{Arneodo:1996qe} and 
      $R_{\rm ABKM}$ the QCD computation to the order indicated.
      The dashed line corresponds to the result of~\cite{Alekhin:2009ni}.
    }
\end{figure}

There exist two possibilities for including fixed-target DIS measurements in global PDF determinations. 
One consists of using the differential cross section $d^2 \sigma/dxdQ^2$ 
(of course with all electro-weak corrections applied, as required by the respective set of experimental data), 
i.e. the left hand side of Eq.~(\ref{eq:sigma}). 
This is the procedure of ABKM~\cite{Alekhin:2009ni}. 
Alternatively, one may work directly with the published values for $F_2$ 
extracted from the data for the cross section using the right hand side of Eq.~(\ref{eq:sigma}).
Although formally equivalent, there are important differences between the two approaches, 
if the latter one does not account on equal terms for the higher-order QCD corrections to $F_2$ and $F_L$.
This may lead to a significant inconsistency in the PDF fit in particular in the case of NMC data analysis.
To clarify this point, let us briefly recall a few essentials. 
NMC was a muon beam experiment at CERN with beam energies of 90, 120, 200, and 280 GeV 
and its data fills the gap in the $(x,Q^2)$-kinematics between the SLAC~\cite{Whitlow:1991uw} 
and the HERA measurements at scales $Q^2 < 10~{\rm GeV}^2$. 
As such it provides a valuable constraint on the gluon PDF at $x > 0.001$.

The extraction of $R$ (or $F_L$) needs at least two cross section measurements 
at different beam energies for a given $x$ and $Q^2$ 
in order to determine the longitudinal component of the cross section 
from the dependence on $y$ in Eq.~(\ref{eq:sigma}).
For NMC, this is not possible in the full kinematic range, 
because the sensitivity to $R$ is substantial only at large $y$, which implies small $x$. 
Thus, only for $x < 0.12$ NMC has extracted a value for $R$ from its data, 
while for $x > 0.12$ almost all NMC data are at $y < 0.40$ with little sensitivity to $R$. 
In this region, $x > 0.12$, NMC has taken $R_{1990}$ from \cite{Whitlow:1991uw}, 
which is based on an empirical parameterization of the SLAC data 
motivated by QCD and including higher-twist terms at large $x$ 
(see also~\cite{Alekhin:2007fh,Blumlein:2008kz}).
A second important issue is concerned with the accuracy of QCD perturbation theory. 
As a matter of fact, the values of $R$ determined by NMC rely on leading-order (LO) QCD predictions only.
However, since several years two more orders in perturbation theory are known for $F_2$ and $F_L$~\cite{Moch:2004xu,Vermaseren:2005qc} 
and these higher-order Wilson coefficients contain the non-trivial $Q^2$ dependence 
and induce big corrections in particular at small $Q^2$.

In Fig.~\ref{fig:R-nmc} we display results for $R$ as published by NMC 
in comparison to values calculated with the ABKM PDFs~\cite{Alekhin:2009ni} in various orders of perturbation theory.
The two $x$ values, $x=0.11$ and $x=0.14$, are chosen to illustrate the different 
analysis strategies of NMC, i.e. either an extraction of $R$ from its data ($x=0.11$)
or the use of $R_{1990}$ from \cite{Whitlow:1991uw} ($x=0.14$).
We find that the value of $R$ obtained with the ABKM~\cite{Alekhin:2009ni} 
fit is in good agreement with $R_{1990}$ at $x > 0.12$. 
This is because similar sets of data (SLAC~\cite{Whitlow:1991uw}) are used in both fits.
Fig.~\ref{fig:R-nmc} also illustrates the impact of the higher-order Wilson coefficients, 
i.e. $F_L^{(1)}$ and $F_L^{(2)}$ from Eq.~(\ref{eq:F2FL}), 
which lead to good perturbative stability of the NNLO prediction even at small values of $Q^2$.
However, the value of $R_{\rm NMC}$ at $x < 0.12$ is quite different from $R_{1990}$ at $x > 0.12$. 
In contrast to the $R$ values computed with ABKM~\cite{Alekhin:2009ni}, 
$R_{\rm NMC}$ does not depend on $Q^2$  at $x < 0.12$, see Fig.~\ref{fig:R-nmc} (left), 
due to assumptions made in the NMC analysis. 
Thus, it is obvious that high precision PDF fits to NLO or NNLO in QCD 
need to be based on the NMC data for the differential cross sections $d^2 \sigma/dxdQ^2$ 
rather than on the NMC results for $F_2$ and $R$, because the latter approach is simply inconsistent.
Of course, a similar statement also holds for the analysis of other fixed-target data, 
where modern parametrization of $R$ with higher-order QCD corrections have to be applied as well 
and, e.g. in the case of BCDMS~\cite{Benvenuti:1989fm,Benvenuti:1989rh}, 
have an impact on the valence quark PDFs~\cite{Blumlein:2006be}.

\bigskip

It is interesting to investigate the consequences of the two alternative treatments 
of the NMC data, which covers the range of $x \sim 0.001 \dots 0.1$, 
and thus is of great importance for the Higgs boson production at current hadron colliders.
For a quantitative analysis we perform a variant of the ABKM fit~\cite{Alekhin:2009ni} 
with the NMC data for the cross section replaced by the data for $F_2$. 
The results for the values of the strong coupling constant $\alphas(M_Z)$ are presented in Tab.~\ref{tab:asvalues}. 
Interestingly, the (inconsistent) NNLO variant of the ABKM fit based on the NMC data on $F_2$ and $R$ 
yields a value of $\alphas(M_Z)=0.1170$, bigger than the default ABKM value by $+0.0035$ and 
rather close to to MSTW~\cite{Martin:2009iq} and the present world average of
$\alphas(M_Z)$ \cite{Bethke:2009jm}~\footnote{%
Note that values of $\alphas(M_Z)$ from NLO, NNLO and N$^3$LO determinations contribute to this average.
In world analyses of DIS and other hard scattering data the extracted NNLO values for $\alphas(M_Z)$ 
are systematically lower than the corresponding NLO ones, e.g. by 
$-0.0044$~\cite{Alekhin:2009ni}, $-0.0031$~\cite{Martin:2009iq} or $-0.0014$~\cite{Blumlein:2006be}.
}.
The resulting shift corresponds to more than $+2\sigma$ standard deviations.
In the NLO case, the shift is smaller, $+0.0009$, whereas it becomes 
even larger if the ${\cal O}(\alphas^3)$ corrections for the Wilson coefficients of $F_L$ are included.
Then, the difference between the two treatments amounts to $+3.6 \sigma$.
This is to be expected, because the alternative treatments of the NMC data are almost equivalent at LO 
(see Fig.~\ref{fig:R-nmc}) and deviate more and more as we include higher and higher orders for $F_L$.
The $\alphas(M_Z)$ values resulting from a consistent
treatment of the NMC data (left column in Tab.~\ref{tab:asvalues}) are in full agreement with other 
recent high precision determinations~\cite{Gehrmann:2009eh,Abbate:2010xh}.
It should also be mentioned here, that the values of $\chi^2$ in all variants of the fit are very similar.
The variations are roughly $\pm 10$ units, which is statistically insignificant given the large number of data points in the fit.
This means that the variation in the ansatz is fully compensated by the changes
in the PDFs and the value of $\alphas(M_Z)$.

\begin{table}[t!]
\centering
{\small
\begin{tabular}{l|ccc}
$\alphas(M_Z)$
  & $\alphas(M_Z)$ with $\sigma_{\rm NMC}$
  & $\alphas(M_Z)$ with $F_2^{\rm NMC}$
  & difference
\\[1ex]
\hline  & & & 
\\[-2ex]
NLO & 
     0.1179(16) &  0.1195(17) &  +0.0026 $\simeq$ 1.5$\sigma$ 
\\
NNLO 
    & 
     {\bf 0.1135(14)} &  0.1170(15) &  +0.0035 $\simeq$ 2.3$\sigma$ 
\\
NNLO +$F_L$ at ${\cal O}(\alphas^3)$& 
     0.1122(14) &  0.1171(14) &  +0.0050 $\simeq$ 3.6$\sigma$ 
\end{tabular}
}
\caption{\small
\label{tab:asvalues}
The values of the strong coupling $\alphas(M_Z)$ obtained 
in global fits of PDFs from variants of the ABKM fit~\cite{Alekhin:2009ni}.
The order of perturbation theory is indicated in the left column 
and in the two central ones the treatment of the NMC data~\cite{Arneodo:1996qe}, 
i.e. a fit to the measured cross sections or to the DIS structure function. 
The right column gives the absolute difference and the relative one in terms of
standard deviations.
The value in bold corresponds to the published result in~\cite{Alekhin:2009ni}.
}
\end{table}

\bigskip

\begin{figure}[t!]
\centering
\vspace*{10mm}
    {
    \includegraphics[angle=0,width=16.0cm]{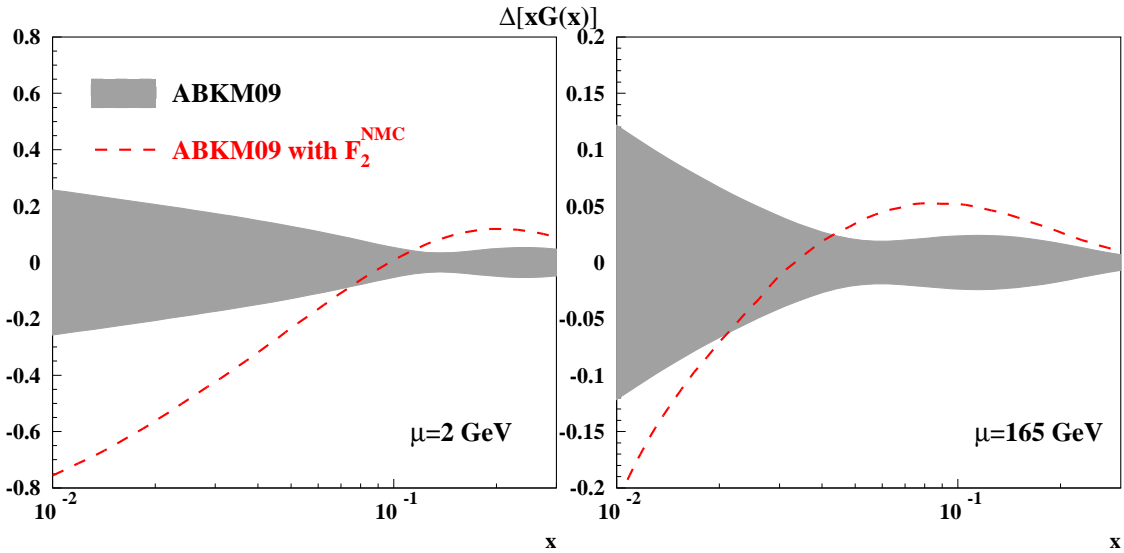}
    }
    \caption{ \small
      \label{fig:gpdf-nmc}
      The absolute uncertainty of the gluon PDF as a function of $x$ at the
      scales $\mu = 2$~GeV (left) and $\mu = 165$~GeV (right) for the ABKM fit of Ref.~\cite{Alekhin:2009ni} (shaded area)
      compared to the difference with the variant of ABKM fit with $F_2^{\rm NMC}$ used (dashes).
    }
\end{figure}
\begin{figure}[t!]
\centering
\vspace*{10mm}
    {
    \includegraphics[angle=0,width=9.0cm]{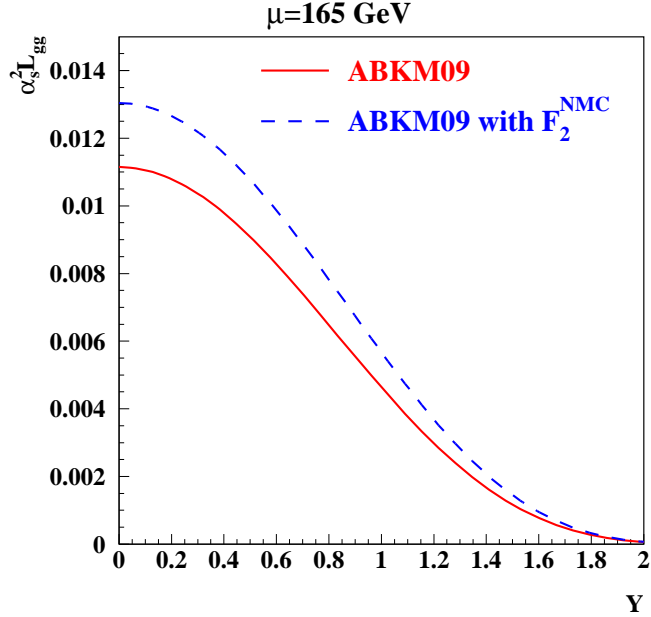}
    }
    \caption{ \small
      \label{fig:Lgg-nmc}
      The gluon luminosity $L_{gg} = g \otimes g$ (weighted by a factor $\alphas^2$) 
      at the scale $\mu=165$~GeV as a function of the Higgs boson's rapidity $Y$.
      The solid line denotes the result of Ref.~\cite{Alekhin:2009ni} 
      and the dashed line the variant with a the fit to $F_2^{\rm NMC}$.
    }
\end{figure}

In Fig.~\ref{fig:gpdf-nmc} we plot the change in the gluon PDF $G(x)$ due to the choice of the NMC data representation.
The variant with a fit to $F_2^{\rm NMC}$ displays significant deviations. 
At the initial scale $\mu=2$~GeV (Fig.~\ref{fig:gpdf-nmc}, left) 
it effectively leads to a larger gluon in the range $x > 0.1$.
One should keep in mind here, that the gluon PDF at larger scales is actually sensitive to all values of $x$ larger than $0.1$, 
because the physical observables emerge as convolutions with the respective Wilson coefficients.
Due to the QCD evolution from $\mu=2$~GeV to the scale $\mu=165$~GeV 
this excess of the gluon PDF then extends to even smaller $x > 0.05$ (Fig.~\ref{fig:gpdf-nmc}, right), 
which is precisely the range in $x$ relevant for Higgs production at Tevatron and the LHC. 
Thus, in the inconsistent variant of the fit to NMC data, one obtains both, 
a larger value of $\alphas(M_Z)$ and a larger gluon PDF. 
This matches with the observed differences between the PDF sets currently available at NNLO in QCD.
For the gluon PDF in the relevant $x$ range, $x \simeq 0.1$, 
(see e.g. Fig.~2 in Ref.~\cite{Alekhin:2010dd}), 
we find good agreement between ABKM and the results of HERAPDF~\cite{herapdf:2009wt}. 
The latter are obtained from a fit without NMC data. 
On the other hand, no agreement exists with JR~\cite{JimenezDelgado:2008hf} and MSTW~\cite{Martin:2009iq}. 
These fits both use the NMC results for $F_2$.
Remarkably, in comparison to ABKM, both the gluon PDF and the $\alphas$ value of MSTW are larger.
In this context it should also be stressed the initial conditions for the gluon PDF 
are significantly correlated with the value of $\alphas(M_Z)$ which determines the speed of the QCD evolution.
Especially at large $x$ a strong anti-correlation is observed (see e.g. Tab.~2 in Ref.~\cite{Alekhin:2009ni}), 
so that a smaller $\alphas(M_Z)$ value implies a larger gluon PDF and vice versa.

\bigskip

Finally, we would like to summarize the impact of the different variants to treat the NMC data 
on the predicted Higgs boson cross sections at the Tevatron and the LHC.
We focus on the dominant channel through gluon-gluon fusion 
and illustrate the cumulative effect of a larger gluon PDF and a larger $\alphas$ value. 
The Born contribution in gluon-gluon fusion is proportional to $\alphas^2$
and the gluon luminosity $L_{gg} = G \otimes G$. 
In Fig.~\ref{fig:Lgg-nmc} we plot the product $\alphas^2 L_{gg}$ at the scale $\mu=165$~GeV as a function of the Higgs boson's rapidity $Y$.
The difference between the ABKM prediction and the (inconsistent) variant with a fit 
to $F_2^{\rm NMC}$ data amounts to an increase of roughly 20\% at central rapidities. 
In order to quantify this enhancement for total cross section predictions we present 
in Tabs.~\ref{tab:hxsvalues-tev} and \ref{tab:hxsvalues-lhc7} the respective numbers.
For the Tevatron (Tab.~\ref{tab:hxsvalues-tev}) the NNLO QCD prediction based
on the fit to the $F_2^{\rm NMC}$ data yields a cross section value
which is 22\% larger than the one from ABKM~\cite{Alekhin:2009ni}. 
This corresponds to a shift of $+2.3\sigma$ standard deviations of the combined uncertainty on $\alphas$ and the PDFs, 
a difference which still increases slightly if the Wilson coefficients for $F_L$ at ${\cal O}(\alphas^3)$ are included.
At NLO however, the difference is of the order of $+1\sigma$ only, 
which is in line with the previous observations in the determination of $\alphas(M_Z)$, 
cf. Tab.~\ref{tab:asvalues}.
For the LHC (Tab.~\ref{tab:hxsvalues-lhc7}) at $\sqrt{s}=7$~TeV center-of-mass energy the same pattern emerges.
At NNLO the inconsistent treatment of the NMC data in the fit leads to a cross
section which is 9\% larger than the ABKM prediction~\cite{Alekhin:2009ni} 
and the difference amounts to $+2.7\sigma$ standard deviations.

The results in Tabs.~\ref{tab:hxsvalues-tev} and \ref{tab:hxsvalues-lhc7} provide a potential explanation 
for the significant spread in the predicted  Higgs cross sections, 
especially between the ABKM and MSTW PDF sets, where the differences are largest.
In the present exclusion region for Higgs masses around $M_H = 165$ GeV, the MSTW prediction at NNLO in QCD 
for Tevatron is +35\% higher than the one of ABKM, 
i.e. a $+4.0 \sigma$ deviation in the combined $\alphas$ and PDF uncertainty.
At the LHC with $\sqrt{s}=7$~TeV the respective MSTW prediction is still +12\% higher, 
which corresponds to a deviation of $+3.6 \sigma$ 
(see the detailed study in~\cite{Alekhin:2010dd} for numbers).
The different handling of the NMC data in the global fits 
that is characterized by either accounting for or neglecting higher-order corrections to $F_L$ 
accounts for the bulk of the observed deviations.

\begin{table}[t!]
\centering
{\small
\begin{tabular}{l|ccc}
$\sigma(H)$
  & $\sigma(H)$ with $\sigma_{\rm NMC}$
  & $\sigma(H)$ with $F_2^{\rm NMC}$
  & difference
\\[1ex]
\hline  & & & 
\\[-2ex]
NLO & 
    0.206(17)~pb & 0.225(18)~pb &  0.019~pb $\simeq$ 1.1$\sigma$ 
\\
NNLO 
    & 
    {\bf 0.253(22)~pb} & 0.309(24)~pb &  0.056~pb $\simeq$ 2.3$\sigma$ 
\\
NNLO +$F_L$ at ${\cal O}(\alphas^3)$& 
    0.242(22)~pb & 0.310(24)~pb &  0.068~pb $\simeq$ 2.8$\sigma$ 
\end{tabular}
}
\caption{\small
\label{tab:hxsvalues-tev}
The predicted cross sections for Higgs boson production in gluon-gluon fusion with $M_H = 165$~GeV 
at Tevatron ($\sqrt{s}=1.96$~TeV) obtained with the PDFs from variants of the ABKM fit~\cite{Alekhin:2009ni}.
The order of perturbation theory is indicated in the left column  
and in the two central ones the treatment of the NMC data~\cite{Arneodo:1996qe}, 
i.e. a fit to the measured cross sections or to the DIS structure function. 
The right column gives the absolute difference and the relative one in terms of
standard deviations.
The value in bold corresponds to the published result~\cite{Alekhin:2010dd}.
}
\end{table}
\begin{table}[t!]
\centering
{\small
\begin{tabular}{l|ccc}
$\sigma(H)$
  & $\sigma(H)$ with $\sigma_{\rm NMC}$
  & $\sigma(H)$ with $F_2^{\rm NMC}$
  & difference
\\[1ex]
\hline  & & & 
\\[-2ex]
NLO & 
    5.73(17)~pb & 5.95(18)~pb &  0.18~pb $\simeq$ 1.0$\sigma$ 
\\
NNLO 
    & 
    {\bf 7.05(23)~pb} & 7.70(23)~pb &  0.65~pb $\simeq$ 2.7$\sigma$ 
\\
NNLO +$F_L$ at ${\cal O}(\alphas^3)$& 
    6.84(21)~pb & 7.68(23)~pb &  0.84~pb $\simeq$ 3.7$\sigma$ 
\end{tabular}
}
\caption{\small
\label{tab:hxsvalues-lhc7}
Same as Tab.~\ref{tab:hxsvalues-tev} for the LHC ($\sqrt{s}=7$~TeV).
}
\end{table}

\bigskip

In summary, we have highlighted the importance of fixed-target DIS data for predictions of
rates for SM Higgs boson production at hadron colliders. 
The use of the NMC data in global PDF analyses allows for different choices.
It is preferable to rely on the differential cross sections $d^2 \sigma/dxdQ^2$ from NMC 
as we have shown that a direct fit to $F_2$ from NMC leads to inconsistencies at higher orders in QCD.
We have illustrated the implications of these options for the determination 
of $\alphas(M_Z)$ and PDFs in a global fits 
and we have computed the rates for Higgs boson production at Tevatron and LHC 
with the results of these fits.
The observed differences in the Higgs cross section allow us to understand 
the deviations in the predictions between the currently available NNLO PDF sets, 
most prominently between ABKM and MSTW, which differ by roughly $4\sigma$ 
in the combined uncertainty for the PDF parameters and $\alphas$.
The details of the DIS fixed-target data analysis are therefore an important issue for the interpretation of the Tevatron data 
and for limits on the mass of a SM Higgs boson~\cite{higgs@tev:2010ar}. 
The current range of excluded Higgs boson masses appears to be much too large. 
It could easily be overestimated by a factor of two based on the reduced rate
for the Higgs boson signal alone~\footnote{%
It has even be argued very recently that the entire mass range would reopen, 
if also the PDF effects for background estimates are taken into account~\cite{Baglio:2011wn}.} 
and this topic needs urgently further investigation.
Potentially, studies of the projected sensitivities for SM Higgs production at the LHC are also affected by this concern. 
In any case, it will be mandatory to base upcoming SM Higgs searches at the LHC on parton luminosities 
from consistent global PDF fits.

\subsection*{Acknowledgments}
We are thankful to A.~Martin and P.~Jimenez-Delgado for discussions. 
This work has been supported by Helmholtz Gemeinschaft under contract VH-HA-101 ({\it Alliance Physics at the Tera\-scale}), 
by the Deutsche Forschungsgemeinschaft in Sonderforschungs\-be\-reich/Transregio~9 
and by the European Commission through contract PITN-GA-2010-264564 ({\it LHCPhenoNet}). 
S.A. also acknowledges partial support from the Russian Foundation for Basic Research under contract RFFI 08-02-91024 CERN\_a.

{\small

}

\end{document}